# Probing Multivariate Indicators for Academic Evaluation

*Journal of Library Science in China* (in press)

Helen F. Xue[1], Loet Leydesdorff[2], Fred Y. Ye [3*]

[1] *Library, Zhejiang University, Hangzhou 310027, CHINA*

[2] *Amsterdam School of Communication Research (ASCoR), University of Amsterdam,*

*PO Box 15793, 1001 NG Amsterdam, The Netherlands*

[3] *School of Information Management, Nanjing University, Nanjing 210023, CHINA*

**Abstract:** *We combine the Integrated Impact Indicator (I3) and the h-index into the I3-type framework and introduce the publication vector $X = (X_1, X_2, X_3)$ and the citation vector $Y = (Y_1, Y_2, Y_3)$, the publication score $I3X=X_1+X_2+X_3$ and the citation score $I3Y=Y_1+Y_2+Y_3$, and alternative indicators based on percentile classes generated by the h-index. These multivariate indicators can be used for academic evaluation. The empirical studies show that the h-core distribution is suitable to evaluate scholars, the $X_1$ and $Y_1$ are applied to measure core impact power of universities, and I3X and I3Y are alternatives of journal impact factor (JIF). The multivariate indicators provide a multidimensional view of academic evaluation with using the advantages of both the h-index and I3.*

---

[*] Corresponding author: Fred Y. Ye, Email: yye@nju.edu.cn

# 1. Introduction

Academic evaluation has continued to be an issue in the academic world, as it is difficult to select and set universal evaluating principles in various complicated situations. However, publications and citations remain the main focuses of academic evaluation, particularly for fundamental research. Citations cannot directly be compared with publications and thus one needs a model or at least a formula. A model can be improved and thus the measurement be refined. Since all models also generate error, the quality of a model depends on the quality of the arguments used for constructing the model.

Since Garfield introduced the journal impact factor (JIF) and set up citation analysis (Garfield, 1955, 1979), these scientometric indicators have been applied to academic evaluations. Hirsch (2005) proposed the h-index, which was rapidly accepted by the scientific community. This promoted the development of quantitative academic indicators. However, both JIF and h-index have their advantages and disadvantages. JIF is basically designed for journals and the h-index for the evaluation of individual scholars.

After developing a set of criteria for an indicator in Leydesdorff et al. (2011), these authors proposed the Integrated Impact Indicator $I3$ (Leydesdorff & Bornmann 2011). $I3$ is based on (*i*) transformation of the citation distribution into a distribution of quantiles and (*ii*) integration (instead of averaging) of the quantile values. (Quantiles are the continuous equivalent of percentiles.) The use of percentiles was recently recommended in the Leiden Manifesto ("Ten principles to guide research evaluation"; Hicks et al., 2015), because average citation rates are heavily dependent on the few highly cited papers in a publication set and bibliometric distributions are very skewed. $I3$ combines citation impact and publication output into a single number – similar to the *h*-index.

The quantile values which are conveniently normalized between zero and hundred provide the weights for the papers, as follows:

$$I3(i) = \sum_{i=1}^{C} f(X_i) \cdot X_i \tag{1}$$

where $X_i$ indicates the percentile ranks and $f(X_i)$ denotes the frequencies of the ranks with $i=[1,C]$ as the percentile rank classes, which means that the measures $X_i$ are divided into C classes each with a scoring function $f(X_i)$ or weight ($w_i$). One can also re-write Eq. (1) as follows:

$$I3(i) = \sum_i w_i X_i ; \sum_i w_i = 1 \tag{2}$$

As an alternative to quantiles, the *h* value of a document set can be used to provide a rank class structure. This combines the advantages of *I3* and *h* into a single framework (Rousseau & Ye, 2012; Ye & Leydesdorff, 2014), which can be applied to academic evaluations based on publications and citations at both group and individual levels. In this study, we elaborate this methodology which was previously applied to journals (Ye et al., 2017), to universities as well as individual scholars.

## 2. Methodology

In many cases, single numbers are used as indicators in academic evaluations. However, a single number can only reflect one side of the overall information and can therefore be expected to have limitations and disadvantages. Possible solutions are multivariate indicators which reflect the multidimensional information. The h-based I3-type multivariate indicators provided a framework of such an elaborate methodology (Ye & Leydesdorff, 2014; Ye et al., 2017).

*3.1 Methods*

Let us assume that the y-axis denotes citations and the x-axis indicates ranked publications from high citation to low citation, then we obtain a publication-citation distribution as in Figure 1. The *h*-index allows us to define three rank classes of both publications and citations in Figure 1. The three classes of publications along the x-axis are: (i) publications in the *h*-core (Ye & Rousseau, 2010; Chen *et al*., 2013) $P_c$, (ii) publications in the *h*-tail $P_t$, (iii) and publications without citations $P_z$. Along the y-axis of the citations one can analogously distinguish among (i) the "excess citations" in the *h*-core (Zhang, 2009, 2013) $C_e=e^2$, (ii) citations to publications in the *h* square of the *h*-core $C_c=h^2$, and (iii) citations to publications in the *h*-tail $C_t=t^2$.

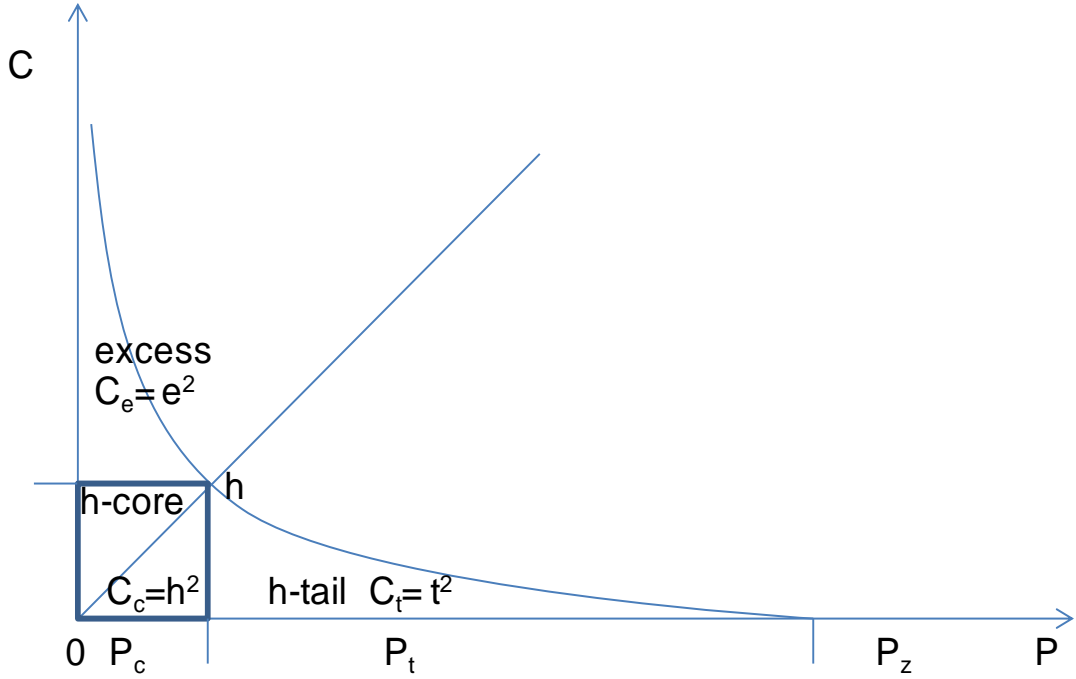

**Fig. 1** The rank distribution of citations versus publications.

Let $x_c=P_c/(P_c+P_t+P_z)$, $x_t=P_t/(P_c+P_t+P_z)$, $x_z=P_z/(P_c+P_t+P_z)$, $y_c=C_c/(C_c+C_t+C_e)$, $y_t=C_t/(C_c+C_t+C_e)$ and $y_e=C_e/(C_c+C_t+C_e)$, we may define two independent vectors as publication vector and citation vector respectively:

$$\mathbf{X} = (X_1, X_2, X_3) = (x_c P_c, x_t P_t, x_z P_z) = (P_c^2/P, P_t^2/P, P_z^2/P) \qquad (3)$$

$$\mathbf{Y} = (Y_1, Y_2, Y_3) = (y_c C_c, y_t C_t, y_e C_e) = (C_c^2/C, C_t^2/C, C_e^2/C) \qquad (4)$$

as well as an *I3*-type publication indicator *I3X* and an *I3*-type citation indicator *I3Y* as follows

$$I3X = x_c P_c + x_t P_t + x_z P_z = X_1 + X_2 + X_3 \qquad (5)$$

$$I3Y = y_c C_c + y_t C_t + y_e C_e = Y_1 + Y_2 + Y_3 \qquad (6)$$

The vector **X** and the score *I3X* represent the relative frequencies of the publications, while the vector **Y** and the score *I3Y* denote the relative frequencies of the citations. For convenient application, citation score in h-core can be merged into $Y_h=Y_1+Y_3=y_h C_h$, where $y_h=C_h/C$, $C_h=C_e+C_c$.

Thus, the h-based I3-type multivariate indicators provide multidimensional indicators: $X_1$ measures publication score in the h-core ($X_1$ and $Y_1$ combination may measure core impact power), $X_2$ measures publication score in h-tail, $Y_h$ measures citation score in h-core, $Y_2$ measures citation score in h-tail, I3X does total publication score, and I3Y does total citation score.

*3.2 Data*

Since $P=P_c+P_t+P_z$, $C=C_h+C_t=C_c+C_t+C_e$, $C_h=C_c+C_e$, $P_c=h$, $C_c=h^2$, one needs to measure only five independent numbers, P, C, $P_z$, $C_h$, h, for the computation of X and Y, *I3X* and *I3Y*, via $P_t=P-P_c-P_z$, $C_c=h^2$, $C_t=C-C_h$, and $C_e=C_h-C_c$. These five values can be obtained easily from bibliometric databases, like by searching Web of Science (WoS) or Scopus.

In order to show the general applicability of these measures, we provide three examples at different levels: 1) individual scholars, we choose the profiles of ourselves in order to avoid issues concerning personal privacy, using 10 years of data from WoS 2005-2015; 2) universities: we chose 25 famous universities, including nine in the USA, nine in China, two in the UK and Germany respectively, and single ones from Australia, Canada, and Japan, with five year data from 2011 to 2015 in WoS; 3) journals, we chose journal datasets 2011- 2015, in the field of electrochemistry (EC). The parameters computed from the datasets are listed in the appendix.

We also collected 2009-2013 data of 25 famous universities and the journal data 2011-2015 in the field of history of the social sciences (HSS), for comparative applications.

**3. Results**

The publication vector $\mathbf{X} = (X_1, X_2, X_3)$ and the citation vector $\mathbf{Y} = (Y_1, Y_2, Y_3)$ are represented by distributed numbers, which are listed in the appendix. The distributed numbers reflect multidimensional academic information, so that the multivariate vectors $\mathbf{X}$ and $\mathbf{Y}$ contribute possible applications as multidimensional indicators. If we want to compare research objects to one another, we can inspect the tabled values of publication vector $\mathbf{X}$ and citation vector $\mathbf{Y}$, where $(X_1, X_2, X_3)$ and/or $(Y_1, Y_2, Y_3)$ rank accordingly. However, if we merge the same-type numbers into one indicator, $I3$-type indicators can be a good choice. $I3\mathrm{X}=X_1+X_2+X_3$ and $I3\mathrm{Y}=Y_1+Y_2+Y_3$ sum the scores of vector $\mathbf{X}$ and $\mathbf{Y}$, respectively. All scores can be plotted into figures.

*3.1 Individual level: scholars*

The scholars' data can be searched via definite field and time span in definite database. Individual dataset is small, so that all indicators can be easily calculated, such as h-index, $X_i$, $Y_i$, I3X, I3Y, even h-core and h-tail distributions of publications and citations. Figure 2 shows the h-core distributions of Leydesdorff L and Ye FY.

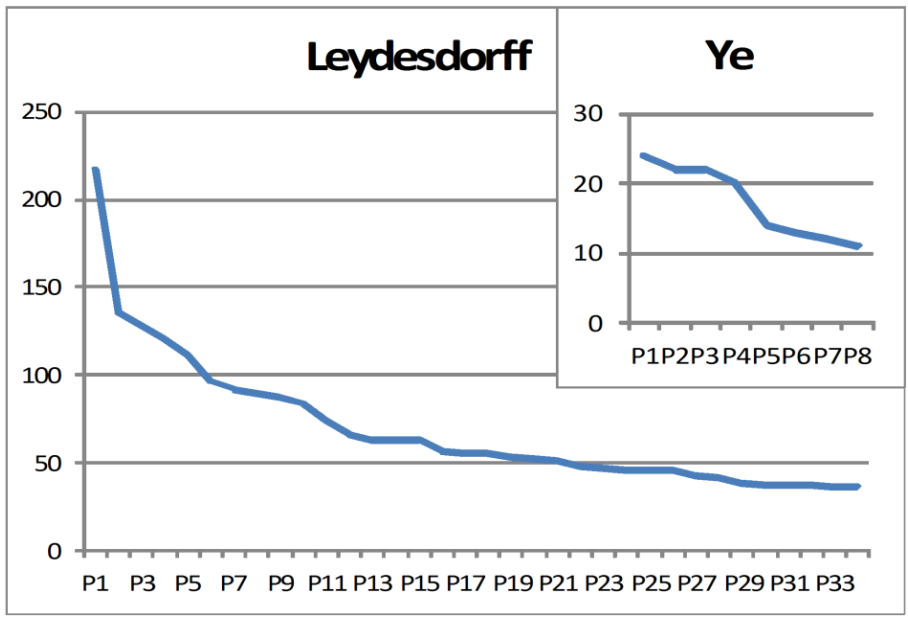

**Fig. 2** Leydesdorff's and Ye's citation-publication distribution in h-core

The indicators of an individual scholar are derived from his/her publications in his/her respective h-core. The multivariate indicators supply a feasible way for mining the indicators. For younger scholars with a lower h-index, the indicators $X_2$ and $Y_2$ can be used to indicate their potential.

*3.2 Group level: universities*

For any university, there are lots of publications and citations distributed in many fields, so that the multivariate indicators provide useful indicators from different perspectives. When we are concerned with the core impact, the h-index, $X_1$ and $Y_1$ provide important h-core information, while ignoring the h-tail. Figure 3 shows the impact of 25 famous universities.

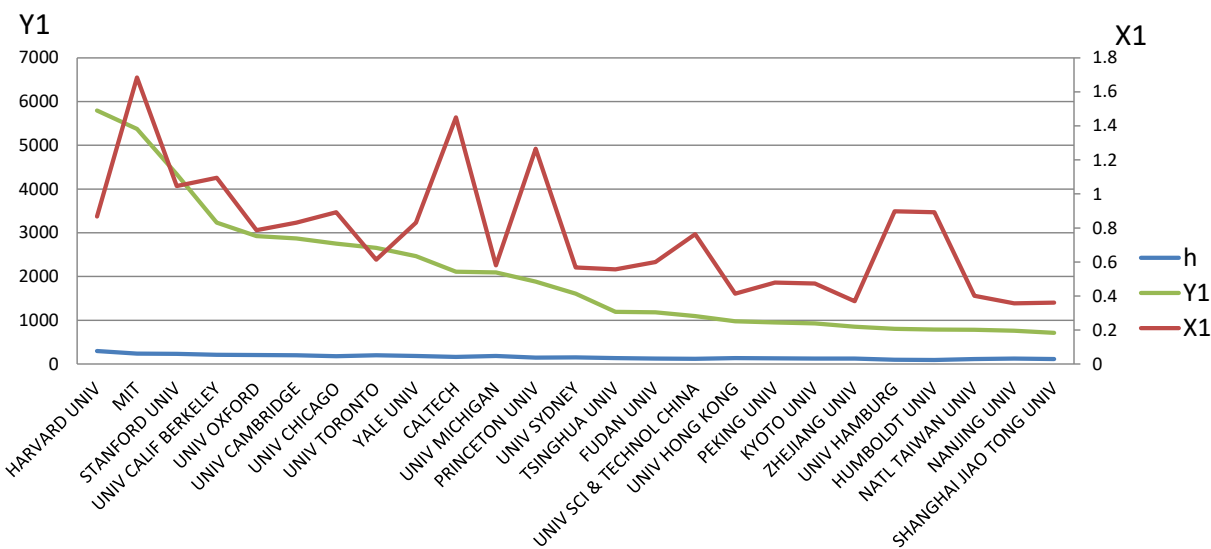

**Fig. 3** The core impact power of 25 famous universities (2011-2015)

Figure 3 shows that Harvard occupies the top-1 position in terms of impact of citations and MIT the top-1 in impact power of publications, while Stanford, Berkeley, Cambridge, Oxford follow these top performers. Among these top universities, Yale and Michigan have core advantages of publications indicated by obvious peaks.

*3.3 Group level: journals*

As all publications and citations are valuable for evaluating in journals, it is recommendable to use I3X and I3Y, which can cover the distribution of publication scores while integrating citation scores of h-core and h-tail. Figure 4 shows this for journals in electro-chemistry (EC).

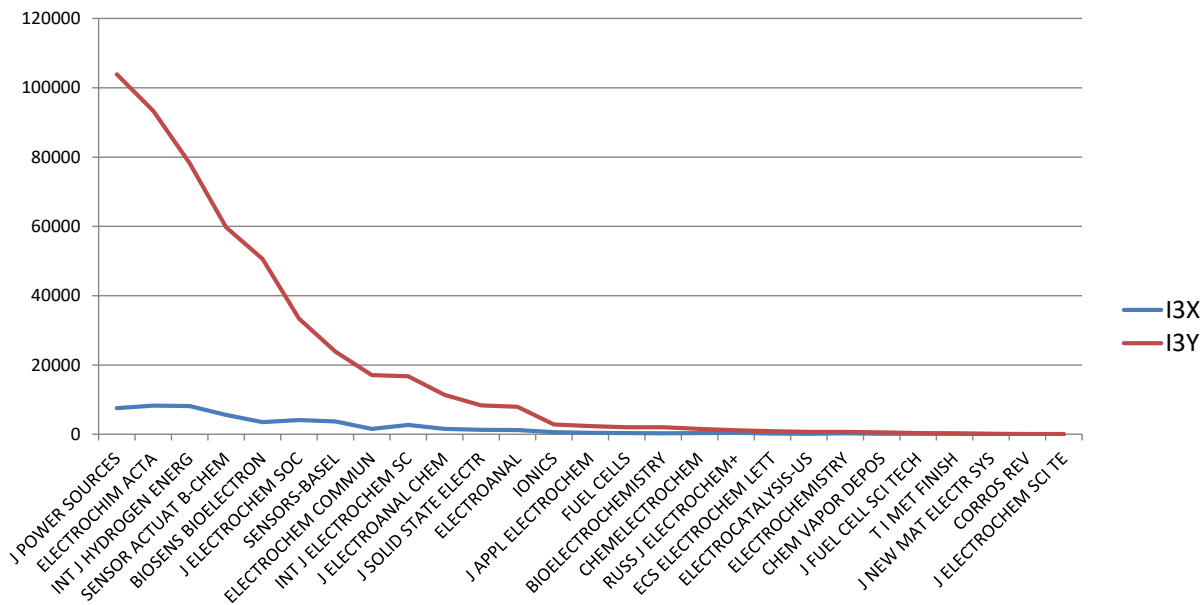

**Fig. 4** The I3X and I3Y of 25 EC journals (2011-2015)

In order to understand the relations among all the indicators, Table 1 shows the Spearman correlations between h and $\{X_i\}$, $\{Y_i\}$ (i=1,2,3), IX3, I3Y for 25 famous universities and Table 2 provides Spearman correlations between JIF and $\{X_i\}$, $\{Y_i\}$ (i=1,2,3), IX3, I3Y for 27 EC journals.

**Table 1** The correlations of multivariate indicators for 25 top-ranked universities (2011-2015)

| Correlations | | Spearman (Sig.(2-tailed)) | | | | |
|---|---|---|---|---|---|---|
| | | h | $Y_1$ | $Y_2$ | $Y_3$ | I3Y |
| Spearman (Sig.(2-tailed)) | h | 1 | .958(.000)* | .838(.000)* | .768(.000)* | .843(.000)* |
| | $X_1$ | .514(.009)* | .678(.000)* | .074(.726) | .824(.000)* | .078(.709) |
| | $X_2$ | .630(.001)* | .440(.028)** | .918(.000)* | .159(.447) | .912(.000)* |
| | $X_3$ | .538(.006)* | .405(.044)** | .775(.000)* | .173(.408) | .778(.000)* |
| | I3X | .671(.000)* | .486(.014)** | .945(.000)* | .188(.369) | .942(.000)* |

*correlation is significant at the 0.01 level (2-tailed); **correlation is significant at the 0.05 level (2-tailed)

**Table 2** The correlations of multivariate indicators for 27 EC journals (2011-2015)

| Correlations | Spearman (Sig.(2-tailed)) |
|---|---|

|  |  | JIF | $Y_1$ | $Y_2$ | $Y_3$ | I3Y |
|---|---|---|---|---|---|---|
| Spearman (Sig.(2-tailed)) | JIF | 1 | .887 (.000)* | .746 (.000)* | .777 (.000)* | .761(.000)* |
|  | $X_1$ | .713 (.000)* | .609 (.001)* | .208 (.297) | .593 (.001)* | .233(.242) |
|  | $X_2$ | .730 (.000)* | .844(.000)* | .995 (.000)* | .679 (.000)* | .995(.000)* |
|  | $X_3$ | -.507 (.007)* | -.275 (.165) | .095(.637) | -.217 (.276) | .068(.735) |
|  | I3X | .678(.000)* | .802(.000)* | .988(.000)* | .667(.000)* | .986(.000)* |

*correlation is significant at the 0.01 level (2-tailed)

Table 1 shows that most multivariate indicators (except a few $X_3$, $Y_3$ and I3X) are positively correlated to the h-index at university level, with Spearman coefficients 0.514, 0.671, 0.843 between h-index and $X_1$, I3X, I3Y respectively. Table 2 shows similar results: most multivariate indicators (except $X_3$) are positive correlations to JIF at journal level. Totally, $\{X_i\}$ (i=1,2) and $\{Y_i\}$ (i=1,2,3), I3X and I3Y are suitable to be independent indicators.

## 4. Discussion and Comparison

The advantages of $X_1$ and $Y_1$ are relative robust like h-index, with non-integral changeability, particularly $Y_1$ can characterize core impact power of citations. In Table 3, we compare the data of 25 famous universities during the periods of 2009-2013 and 2011-2015, in terms of h-index and $Y_1$. One can see the quick development of the Chinese universities compared to the world-class universities.

Table 3. The Change of Universities' h-indices and $Y_1$

| 2009-2013 | | | 2011-2015 | | |
|---|---|---|---|---|---|
| UNIV. | h | $Y_1$ | UNIV. | h | $Y_1$ |
| HARVARD | 272 | 4763.45 | HARVARD | 299 | 5794.92 |
| MIT | 217 | 4506.3 | MIT | 241 | 5374.34 |
| UC BERKELEY | 203 | 3426.45 | STANFORD | 231 | 4335.86 |
| STANFORD | 202 | 3242.72 | UC BERKELEY | 210 | 3232.96 |
| CAMBRIDGE | 190 | 2822.44 | OXFORD | 206 | 2926.63 |
| OXFORD | 192 | 2782.86 | CAMBRIDGE | 201 | 2870.43 |
| CHICAGO | 164 | 2387.89 | CHICAGO | 178 | 2754.38 |

| MICHIGAN | 181 | 2166.96 | TORONTO | 200 | 2654.09 |
| --- | --- | --- | --- | --- | --- |
| CALTECH | 154 | 2081.41 | YALE | 183 | 2464.35 |
| TORONTO | 178 | 2051.62 | CALTECH | 161 | 2111.04 |
| YALE | 161 | 1840.59 | MICHIGAN | 186 | 2094.91 |
| PRINCETON | 133 | 1559.91 | PRINCETON | 146 | 1885.78 |
| TSINGHUA | 111 | 878.081 | SYDNEY | 153 | 1608.35 |
| SYDNEY | 120 | 853.671 | TSINGHUA | 135 | 1195.78 |
| PEKING | 112 | 799.809 | FUDAN | 128 | 1183.3 |
| FUDAN | 102 | 734.071 | USTC | 120 | 1098.46 |
| KYOTO | 114 | 714.517 | HONG KONG | 136 | 977.568 |
| HONG KONG | 116 | 700.921 | PEKING | 130 | 949.031 |
| HUMBOLDT | 81 | 609.58 | KYOTO | 126 | 931.677 |
| HAMBURG | 82 | 574.298 | ZHEJIANG | 126 | 851.592 |
| USTC | 89 | 552.153 | HAMBURG | 97 | 805.082 |
| NANJING | 98 | 487.427 | HUMBOLDT | 92 | 789.657 |
| SHANGHAI JIAO TONG | 92 | 459.206 | NATL TAIWAN | 116 | 786.014 |
| ZHEJIANG | 95 | 428.322 | NANJING | 123 | 759.485 |
| NATL TAIWAN | 85 | 294.895 | SHANGHAI JIAO TONG | 116 | 712.446 |

There are disciplinary differences, which could affect the applications of the multivariate indicators. For example, comparing the journals of history of the social sciences with the journals of electrochemistry, the relation of I3X and I3Y as well as their correlations to JIF show differences in Figure 5 and Table 4.

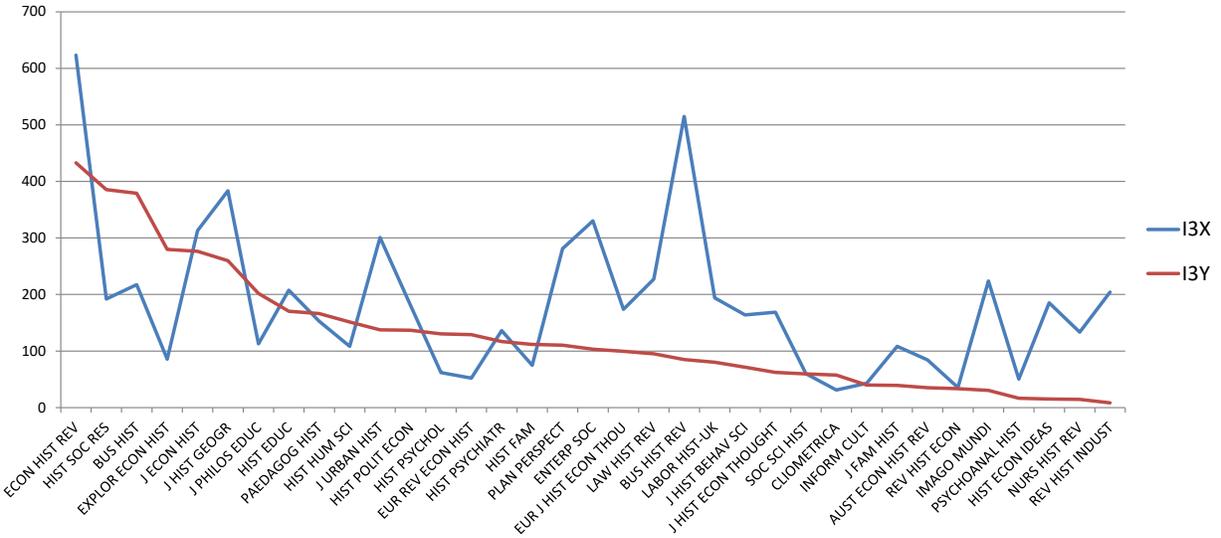

**Fig.5** The I3X and I3Y of 35 HSS journals (2011-2015)

**Table 4** The correlations of multivariate indicators for 35 HSS journals (2011-2015)

| Correlations | | Spearman (Sig.(2-tailed)) | | | | |
| --- | --- | --- | --- | --- | --- | --- |
| | | JIF | $Y_1$ | $Y_2$ | $Y_3$ | I3Y |
| Spearman (Sig.(2-tailed)) | JIF | 1 | .690 (.000)* | .521 (.001)* | .634 (.000)* | .527(.001)* |
| | $X_1$ | .548 (.001)* | .774 (.000)* | .343 (.044)** | .626 (.001)* | .353(.037)** |
| | $X_2$ | .470 (.004)* | .347(.041)** | .880 (.000)* | .408 (.015)** | .876(.000)* |
| | $X_3$ | .006 (.974) | -.037 (.832) | .084 (.632) | -.172(.323) | .088(.614) |
| | I3X | .131(.455) | .080(.647) | .348(.041)** | -.041(.813) | .352(.038)** |

*correlation is significant at the 0.01 level (2-tailed); **correlation is significant at the 0.05 level (2-tailed)

Here we see that the correlations in multivariate indicators are much lower in the social sciences. Particularly, I3X is no longer correlated to JIF; it is an independent indicator. Therefore, the multivariate indicators provide richer measurement information than single indicators.

In general, if we want to compare two academic subject or object A and B, we may compare all elements of their academic matrices $M_A$ and $M_B$. If all elements in $M_A$ are better than $M_B$ (recorded as $\{M_A\} \succ \{M_B\}$, not always A>B; for X3, smaller value is better), we can say A is better than B. More generally, academic tensor T is suggested to be a generalized measure including matrix. We can compare all elements of their academic tensors $T_A$ and $T_B$. If all elements in $T_A$ are better than $T_B$ (recorded as $\{T_A\} \succ \{T_B\}$), we can say A is better than B.

## 5. Conclusions

The multivariate indicators, including publication vector $\mathbf{X} = (X_1, X_2, X_3)$ and citation vector $\mathbf{Y} = (Y_1, Y_2, Y_3)$, publication score $I3X = X_1+X_2+X_3$ and citation score $I3Y = Y_1+Y_2+Y_3$, as well as their elements and integrated indices, provide a methodological framework for extensive academic measurement. Most of them are positively correlated to the h-index and JIF, with relative independence (Spearman coefficients 0.5~0.9), so that they can be considered as independent indicators, which provide multidimensional views for academic evaluation.

Particularly, the core-tail measurements of $\mathbf{X}$ and $\mathbf{Y}$, as well as $I3X$ and $I3Y$ combine the advantages of the *h*-index and *I3*: (i) the publications and not only the citations are appreciated; (ii) the indicators are non-parametric; (iii) the results are easy to obtain from WoS or Scopus data; (iv) the results can be plotted via X-Y system. We note that these indicators do not require reference sets as when using quantile or percentile values (Bornmann et al., 2013); the distributions are generated from the h-classes as shown in Figure 1 above.

We plan to develop further studies with applications and extensions of these multivariate indicators.

# Acknowledgements

We acknowledge the National Natural Science Foundation of China Grant No 71673131 for partly financial supports.

**Appendix**

Table A1. Scholars' data

| Indicator | P | h=P_c | P_z | C | C_h | X1 | X2 | X3 | Y1 | Y2 | Y3 |
|---|---|---|---|---|---|---|---|---|---|---|---|
| Leydesdorff L | 145 | 35 | 15 | 3673 | 2404 | 8.44828 | 62.2414 | 1.551724 | 408.5557 | 438.4321 | 378.4484 |
| Ye FY | 27 | 8 | 4 | 193 | 138 | 2.37037 | 8.33333 | 0.592593 | 21.2228 | 15.67358 | 28.37306 |

Table A2. Publication and citation vectors of 25 famous universities ranked by h-index based on WoS data from 2009 to 2013.

| University (ISI Abbreviated Name) | Univ h-index | Publication Vector | | | Citation Vector | | |
|---|---|---|---|---|---|---|---|
| | | X1 | X2 | X3 | Y1 | Y2 | Y3 |
| HARVARD UNIV | 272 | 1.079165 | 20582.44 | 2591.597 | 3426.448 | 334689.3 | 4480.493 |
| MIT | 217 | 1.392601 | 11333.85 | 522.5067 | 2081.409 | 175299.2 | 3082.672 |
| STANFORD UNIV | 203 | 0.830076 | 19064.02 | 4838.446 | 2822.444 | 320858.5 | 3592.503 |
| UNIV CALIF BERKELEY | 202 | 0.840999 | 10397.27 | 5768.135 | 2387.891 | 175558.4 | 6812.783 |
| UNIV OXFORD | 192 | 0.807298 | 44786.92 | 8136.022 | 4763.454 | 910285.6 | 2386.228 |
| UNIV CAMBRIDGE | 190 | 0.715319 | 4545.997 | 822.7618 | 574.2979 | 48509.93 | 1322.582 |
| UNIV TORONTO | 181 | 0.766562 | 4067.064 | 776.5024 | 609.5801 | 45841.01 | 725.9668 |
| UNIV MICHIGAN | 178 | 0.397346 | 16171.09 | 2814.811 | 714.5166 | 188539.5 | 637.3312 |
| YALE UNIV | 164 | 0.605866 | 22933.74 | 6451.1 | 2166.962 | 355131.9 | 3756.614 |
| UNIV CHICAGO | 161 | 1.52057 | 18113.27 | 1612.713 | 4506.299 | 310350.5 | 5967.65 |
| CALTECH | 154 | 0.787154 | 19806.28 | 5592.797 | 2782.855 | 317658.6 | 6796.675 |
| UNIV SYDNEY | 133 | 1.135803 | 8201.824 | 1099.995 | 1559.91 | 112274.4 | 5373.22 |
| PRINCETON UNIV | 120 | 0.922583 | 20599.48 | 4332.119 | 3242.723 | 373369.8 | 2358.782 |
| UNIV HONG KONG | 116 | 0.418131 | 14795.44 | 4006.171 | 853.6706 | 178022.5 | 1739.573 |
| TSINGHUA UNIV | 114 | 0.552641 | 24793.15 | 6599.81 | 2051.624 | 364049 | 2585.307 |
| PEKING UNIV | 112 | 0.725977 | 15521.04 | 4035.065 | 1840.585 | 269402.6 | 1784.917 |
| FUDAN UNIV | 111 | 0.481684 | 12800.7 | 2125.225 | 878.081 | 127291.7 | 863.3201 |
| KYOTO UNIV | 102 | 0.442985 | 14005.98 | 2426.956 | 799.8086 | 152382.1 | 620.4136 |
| ZHEJIANG UNIV | 98 | 0.485374 | 10470.28 | 1882.337 | 734.0713 | 113237.5 | 416.0963 |
| NANJING UNIV | 95 | 0.342165 | 18748.04 | 3696.568 | 700.9207 | 210327.2 | 536.6424 |
| UNIV SCI & TECHNOL CHINA | 92 | 0.300191 | 15794.25 | 2771.264 | 487.4272 | 154049.6 | 417.7406 |
| SHANGHAI JIAO TONG UNIV | 89 | 0.302189 | 13202.65 | 2694.276 | 459.2057 | 120447.5 | 701.8614 |
| NATL TAIWAN UNIV | 85 | 0.231481 | 14935.64 | 2913.472 | 294.8955 | 145383.4 | 495.7767 |
| UNIV HAMBURG | 82 | 0.537929 | 8464.17 | 818.6611 | 552.1529 | 87193.86 | 335.2364 |

| University (ISI Abbreviated Name) | Univ h-index | Publication Vector | | | Citation Vector | | |
|---|---|---|---|---|---|---|---|
| | | X1 | X2 | X3 | Y1 | Y2 | Y3 |
| HUMBOLDT UNIV | 81 | 0.265184 | 16452.78 | 3102.155 | 428.3223 | 160340.9 | 223.6169 |

Table A2. Publication and citation vectors of 25 famous universities ranked by h-index based on WoS data from 2011 to 2015.

| University (ISI Abbreviated Name) | Univ h-index | Publication Vector | | | Citation Vector | | |
|---|---|---|---|---|---|---|---|
| | | X1 | X2 | X3 | Y1 | Y2 | Y3 |
| HARVARD UNIV | 299 | 1.094619 | 24314.06 | 1913.433 | 3232.96 | 396695.9 | 7903.913 |
| MIT | 241 | 1.450045 | 12384.46 | 449.9271 | 2111.042 | 198999.9 | 5201.47 |
| STANFORD UNIV | 231 | 0.831177 | 23130.99 | 4552.135 | 2870.434 | 394468.6 | 5247.233 |
| UNIV CALIF BERKELEY | 210 | 0.892381 | 12471.7 | 5746.589 | 2754.382 | 214970 | 7670.005 |
| UNIV OXFORD | 206 | 0.867213 | 52946.5 | 8107.924 | 5794.924 | 1042924 | 5936.042 |
| UNIV CAMBRIDGE | 201 | 0.898491 | 5597.244 | 705.9741 | 805.0824 | 59230.25 | 3583.228 |
| UNIV TORONTO | 200 | 0.891792 | 4919.386 | 693.7473 | 789.6574 | 55358.94 | 1429.996 |
| UNIV MICHIGAN | 186 | 0.47295 | 18008.83 | 2335.886 | 931.6765 | 212361.4 | 828.0422 |
| YALE UNIV | 183 | 0.579624 | 26893.7 | 6328.984 | 2094.907 | 430861 | 2882.996 |
| UNIV CHICAGO | 178 | 1.684826 | 21638.54 | 1389.099 | 5374.339 | 384660.6 | 9750.973 |
| CALTECH | 161 | 0.786624 | 24606.31 | 5552.343 | 2926.635 | 413946.2 | 7558.184 |
| UNIV SYDNEY | 153 | 1.26542 | 9514.728 | 968.4489 | 1885.783 | 141185.6 | 5139.476 |
| PRINCETON UNIV | 146 | 1.045679 | 24906.8 | 4496.608 | 4335.864 | 459288.5 | 4464.867 |
| UNIV HONG KONG | 136 | 0.567175 | 18669.88 | 4325.257 | 1608.351 | 236197.7 | 3317.497 |
| TSINGHUA UNIV | 135 | 0.613459 | 29408.69 | 6901.935 | 2654.091 | 439022.9 | 3884.249 |
| PEKING UNIV | 130 | 0.831096 | 18475.52 | 4083.185 | 2464.349 | 319306.9 | 3587.126 |
| FUDAN UNIV | 128 | 0.557271 | 18821.43 | 1840.817 | 1195.784 | 206680.1 | 1431.567 |
| KYOTO UNIV | 126 | 0.478971 | 19432.44 | 2279.871 | 949.0312 | 227797.6 | 1640.276 |
| ZHEJIANG UNIV | 126 | 0.599949 | 14670.63 | 1879.865 | 1183.301 | 168309.7 | 1000.845 |
| NANJING UNIV | 123 | 0.414086 | 23361.4 | 3347.527 | 977.5684 | 282767.1 | 814.6924 |
| UNIV SCI & TECHNOL CHINA | 120 | 0.356455 | 23518.15 | 2710.625 | 759.4846 | 245094.5 | 693.9933 |
| SHANGHAI JIAO TONG UNIV | 116 | 0.361216 | 19821 | 2664.592 | 712.4462 | 195774.3 | 1222.995 |
| NATL TAIWAN UNIV | 116 | 0.401708 | 17416.95 | 2541.646 | 786.0145 | 174981.4 | 1129.731 |
| UNIV HAMBURG | 97 | 0.763764 | 11735.74 | 789.8526 | 1098.462 | 133663.5 | 1277.13 |
| HUMBOLDT UNIV | 92 | 0.368909 | 23381.48 | 2908.594 | 851.592 | 238996.7 | 674.8668 |

Table A3. Publication and citation vectors of 27 journals ranked by JIF in the field of electrochemistry based on WoS data from 2011 to 2015. The journals are ranked by their Journal Impact Factors (JIF) 2015.

| Journal (JCR Abbreviated Title) | JIF | Publication Vector | | | Citation Vector | | |
|---|---|---|---|---|---|---|---|
| | | X1 | X2 | X3 | Y1 | Y2 | Y3 |

| Journal | IF | | | | | | |
|---|---|---|---|---|---|---|---|
| *BIOSENS BIOELECTRON* | 6.395 | 1.356003 | 3534.652 | 1.471358 | 413.0753 | 50069.81 | 120.5079 |
| *J POWER SOURCES* | 5.314 | 0.937729 | 7557.725 | 16.0322 | 489.1992 | 103161.5 | 219.6785 |
| *ELECTROCHEM COMMUN* | 4.417 | 0.490168 | 8274.604 | 47.46639 | 213.9841 | 93083.97 | 32.97644 |
| *ELECTROCHIM ACTA* | 4.119 | 0.571882 | 5554.181 | 16.47085 | 186.1962 | 59566.58 | 37.42595 |
| *SENSOR ACTUAT B-CHEM* | 3.987 | 1.701574 | 1568.095 | 4.106891 | 380.931 | 16589.19 | 152.7751 |
| *CHEMELECTROCHEM* | 3.27 | 1.91687 | 284.3056 | 3.91198 | 160.8205 | 1837.389 | 39.3888 |
| *BIOELECTROCHEMISTRY* | 3.231 | 0.700971 | 334.4175 | 12.73981 | 53.0407 | 1533.366 | 9.778185 |
| *J ELECTROANAL CHEM* | 2.553 | 0.347822 | 8080.94 | 91.32054 | 138.5712 | 78089.58 | 20.19831 |
| *J ELECTROCHEM SOC* | 2.461 | 0.598673 | 4032.372 | 88.23733 | 241.2083 | 33041.08 | 141.447 |
| *INT J HYDROGEN ENERG* | 2.371 | 0.627907 | 1535.078 | 29.79845 | 110.939 | 11231.28 | 42.6959 |
| *ELECTROANAL* | 2.179 | 0.544135 | 1208.825 | 26.66264 | 77.01084 | 7842.054 | 27.3147 |
| *J APPL ELECTROCHEM* | 2.143 | 1.184426 | 159.0533 | 3.688525 | 60.43488 | 603.1614 | 23.44428 |
| *J SOLID STATE ELECTR* | 2.099 | 0.521432 | 1265.181 | 44.07216 | 84.75002 | 8287.328 | 17.20556 |
| *ELECTROCATALYSIS-US* | 2.074 | 0.714919 | 427.5391 | 20.22009 | 56.6383 | 2219.753 | 93.54062 |
| *ECS ELECTROCHEM LETT* | 1.93 | 0.389484 | 612.3165 | 44.59202 | 41.52608 | 2826.369 | 6.075526 |
| *CHEM VAPOR DEPOS* | 1.656 | 0.488881 | 3543.361 | 200.0911 | 228.0533 | 23512.43 | 112.7164 |
| *FUEL CELLS* | 1.648 | 0.585938 | 202.7109 | 21.09375 | 34.88973 | 846.0813 | 9.596141 |
| *IONICS* | 1.627 | 0.945378 | 118.5882 | 12.71008 | 52.78936 | 489.2857 | 2.50365 |
| *SENSORS-BASEL* | 1.571 | 0.552901 | 362.6638 | 19.53754 | 38.17309 | 2013.312 | 1.937818 |
| *INT J ELECTROCHEM SC* | 1.266 | 0.232688 | 2554.359 | 175.3513 | 48.13264 | 16719.86 | 5.17147 |
| *CORROS REV* | 1.05 | 0.719101 | 21.75281 | 15.38202 | 16.06275 | 71.47059 | 12.29804 |
| *ELECTROCHEMISTRY* | 0.714 | 0.243243 | 157.1368 | 127.7449 | 17.57288 | 637.0246 | 24.20424 |
| *J FUEL CELL SCI TECH* | 0.64 | 0.220109 | 97.06793 | 78.53261 | 11.67438 | 349.1975 | 2.569395 |
| *T I MET FINISH* | 0.57 | 0.146312 | 269.3881 | 143.0907 | 9.959864 | 1137.312 | 2.133333 |
| *RUSS J ELECTROCHEM+* | 0.502 | 0.264706 | 65.89542 | 78.51307 | 13.75472 | 273.2096 | 2.568134 |
| *J ELECTROCHEM SCI TE* | 0.462 | 0.297619 | 19.04762 | 18.10714 | 5.482456 | 54.74561 | 0.877193 |
| *J NEW MAT ELECTR SYS* | 0.4 | 0.172249 | 34.56938 | 66.62201 | 5.355372 | 152.3306 | 0.809917 |